\documentstyle[11pt,newpasp,twoside,epsf]{article}
\reversemarginpar

\begin{document}
\title{Low-Ionization BAL QSOs in Ultraluminous Infrared Systems}
 \author{Gabriela Canalizo}
\affil{Institute of Geophysics and Planetary Sciences, Lawrence Livermore 
National Laboratory, 7000 East Avenue, L413, Livermore, CA 94550}
\author{Alan Stockton}
\affil{Institute for Astronomy, University of Hawaii, 2680 Woodlawn
 Drive, Honolulu, HI 96822}

\begin{abstract}
Low-ionization broad absorption line (BAL) QSOs present properties that
cannot generally be explained by simple orientation effects.  We have
conducted a deep spectroscopic and imaging study of the host galaxies
of the only four BAL QSOs that are currently known at $z < 0.4$, 
and found that all four objects reside in dusty, starburst or
post-starburst, merging systems.
The starburst ages derived from modeling the stellar populations are
in every case a few hundred million years or younger.   There is strong
evidence that the ongoing mergers triggered both the starbursts and the
nuclear activity, thus indicating that the QSOs have been recently triggered
or rejuvenated.   The low-ionization BAL phenomenon then appears to be 
directly related to young systems, and it may represent a short-lived 
stage in the early life of a large fraction of QSOs.
\end{abstract}

\section{Introduction}

Broad absorption line (BAL) QSOs comprise
$\sim12$\% of the QSO population in current magnitude-limited samples.
The standard view is that the BAL clouds have a small covering factor as seen 
from the QSO nucleus, implying that essentially all radio-quiet QSOs would 
be classified as BAL QSOs if observed from the proper angle.

An even rarer class comprising only $\sim 1.5\%$ of all radio-quiet 
QSOs in optically-selected samples, the low-ionization BAL (hereafter lo-BAL)
QSOs, pose serious problems to orientation-based models.
Lo-BAL QSOs have considerably different
broad emission line properties, are substantially redder than non-BAL 
QSOs, and are intrinsically X-ray quiet (Weymann et al. 1991; Sprayberry 
\& Foltz 1992; Green 2001).   In addition, the radio properties of lo-BAL 
QSOs may indicate that no preferred viewing orientation is necessary to 
observe 
BAL systems in the spectra of quasars since BALs are present in both 
flat and steep spectrum quasars, {\it i.e.,} objects that are presumably
viewed along the jet axis and at high inclination respectively (Brotherton
2001; Becker et al. 2000; Gregg et al. 2000).
Thus lo-BAL QSOs are thought to constitute a different class of QSOs, 
having more absorbing material and more dust (Voit et al. 1993; 
Huts\'{e}mekers et al. 1998).

\section {A Sign of Youth?}

An intriguing possibility is that the lo-BAL phenomenon may represent a stage 
in the early life of QSOs, either in the form of young QSOs ``in
the act of casting off their cocoons of gas and dust'' (Voit et al.
1993; see also Egami et al. 1996, and Hazard et al. 1984), or as
the result of outflows driven by supermassive starbursts (L\'{\i}pari
1994; Shields 1996).  

We have carried out deep Keck spectroscopic observations of the host 
galaxies of the four lo-BAL QSOs that are currently known at $z < 0.4$: 
PG\,1700+518 (Canalizo \& Stockton 1997), Mrk\,231 and IRAS\,07598+6508 
(Canalizo \& Stockton 2000), and IRAS\,14026+4341 (Canalizo \& Stockton 2001, 
in preparation).   
These objects have nuclear properties that are relatively rare in the 
classical QSO population ({\it e.g.,} Boroson \& Meyers 1992; Turnshek et al. 
1997): (1) strong Fe\,II emission, with the flux ratio 
Fe\,II $\lambda4570$/H$\beta$ $\geq$ 1, and (2) very weak or no [O\,III]
emission (see Table 1).
In addition, we have found the following properties in the low redshift 
sample:
(3) every lo-BAL QSO resides in an ultraluminous infrared galaxy 
(ULIG; {\it i.e.,} log $L_{ir}/L_{\sun} \geq 12$); (4) they have a small 
range in far infrared (FIR) colors, intermediate between those 
characteristic of ULIGs and QSOs;  (5) the host galaxies show signs of 
strong tidal interaction, and they appear to be major mergers (Fig.~1); 
(6) spectra of their host galaxies (Fig.~2) show unambiguous 
interaction-induced star formation, with post-starburst ages $\la 250$ Myr.

\begin{table}
\caption{Properties of low-redshift low-ionization BAL QSOs}
\begin{tabular}{lccccccc}
\tableline
Object&Redshift& log              & REW\tablenotemark{a}      & Fe\,II/  & Tail(s)& Dyn.& Sb. \\
Name  & $z$ & L$_{ir}$/L$_{\sun}$ & [O\,III] & H$\beta$ & Length\tablenotemark{b} & Age\tablenotemark{c} & Age\tablenotemark{c} \\
\tableline
IRAS\,07598+6508 & 0.1483 & 12.41 & 0  & 2.6 & 50 & 160 & 30  \\
Mrk\,231         & 0.0422 & 12.50 & 0  & 2.1 & 35 & 110 & 40  \\
IRAS\,14026+4341 & 0.3233 & 12.77 & 0  & 1.0 & 37 & 120 & TBD \\
PG\,1700+518     & 0.2923 & 12.58 & 2  & 1.4 & 13 & 40  & 85  \\
\tableline
\tableline
\end{tabular}
\begin{footnotesize}

$^{a}$ Rest equivalent width (REW) in \AA \\
$^{b}$ Tail lengths in kpc, assuming $H_{0} = 75$ km\,s$^{-1}$ Mpc$^{-1}$ and $q_{0}=0.5$ \\
$^{c}$ Dynamical and starburst ages in Myr
\end{footnotesize}
\end{table}

\begin{figure}[htb]
\plotone{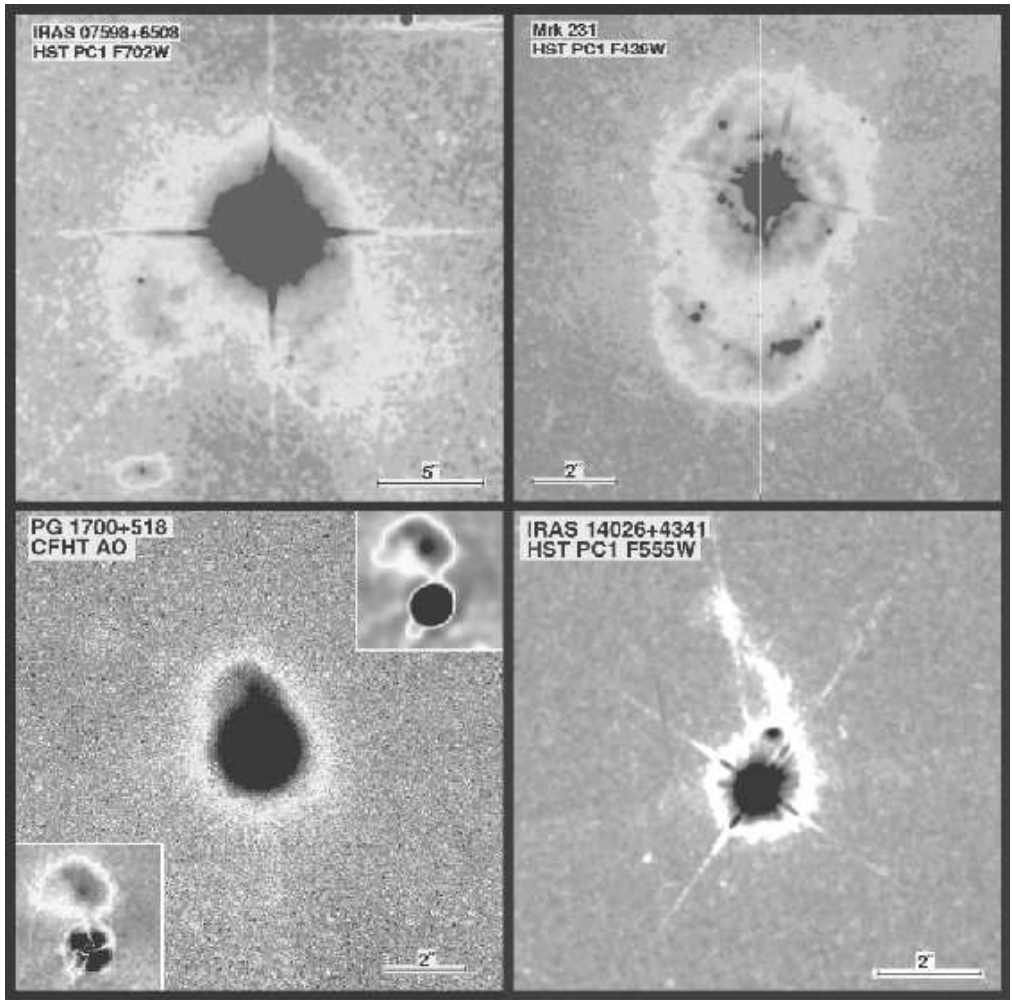}
\caption{Adaptive Optics and HST images of the host galaxies of low-redshift 
lo-BAL QSOs.  Every host galaxy shows signs of recent strong tidal 
interaction.}
\end{figure}

\begin{figure}[hbt]
\plotone{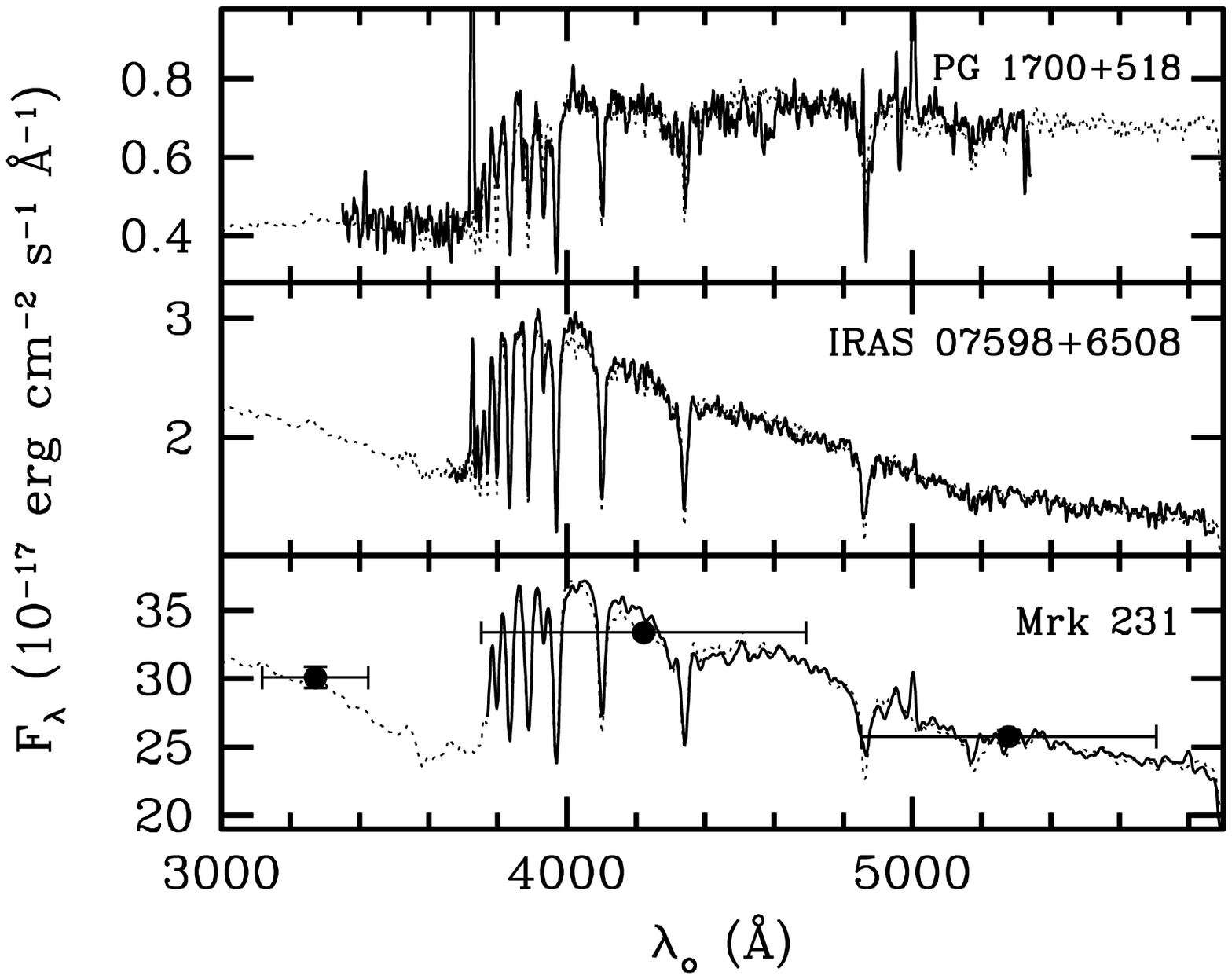}
\caption{Stellar populations in the host galaxies of low-z lo-BAL QSOs.
Panels display Keck LRIS spectra of the host galaxies of 3 of the lo-BAL
QSOs in rest frame (solid trace).   Superposed are the best fitting models 
(dotted traces),
consisting of the sum of an underlying old stellar population model and a
young instantaneous starburst model (Bruzual \& Charlot 1996).   For
details, see Canalizo \& Stockton 1997.  We have recently obtained Keck ESI
spectroscopy of the fourth object, IRAS 14026+4341, which shows a large
H\,II region $\sim2\farcs5$ North of the nucleus, as well as several 
post-starburst regions with populations similar to those of the other
three objects (Canalizo \& Stockton 2001, in preparation).}
\end{figure}

The spatially resolved spectra of the hosts clearly show a concentration of 
material towards the central regions in timescales that are consistent with 
the dynamical age for the tidal interaction (Canalizo \& Stockton 2000; 
Table~1).  Thus, it is clear that there has been a recent flow of 
gas towards the central regions which fueled the centrally concentrated 
starbursts in each of these objects.  Moreover, there is strong evidence 
that these QSOs have been recently fueled (either for the first time, or 
simply rejuvenated).   In agreement with this scenario,
Mathur et al. (2001) find that the X-ray flux of the lo-BAL QSO PHL 5200 
is highly absorbed with a very steep power-law slope of $\alpha = 1.7 \pm 0.4$
(compared to the mean slope for non-BAL QSOs of $\alpha = 0.67 \pm 0.11$).
Such a steep slope may be the result of a high accretion rate close to
the Eddington limit, which may in turn be indicative of a recent fueling of the
black hole ({\it i.e.,} a young QSO).

\section{Proposed Model}

Our results support those interpretations of the lo-BAL phenomenon
which imply young systems.  Here we propose a model that accounts for the
observed properties in the four low-redshift lo-BAL QSOs.  A major
merger between galaxies of similar mass triggers intense bursts of star
formation.  As the gas concentrates in nuclear regions, the QSO activity
is ignited.  Along with the  gas, dust is concentrated in the central
1 or 2 kpc, resulting in a dust-enshrouded QSO.   The dust cocoon shields
the narrow line region from the ionizing radiation coming from the 
central continuum source.   The resulting low ionization parameter
and the dusty environment increases the relative prominence of Fe\,II 
emission.
A lo-BAL phase comes next, consisting of widespread outflows whereby the 
QSO expels the shroud of gas and dust (e.g. Voit, Weymann, \& Korista 1993).  
As the ionizing photons are able to escape through holes on the cocoon 
poked by the outflows, nuclear and extended [O\,III] appear.  A cocoon
with holes may explain the different lightpaths in some BAL QSOs inferred 
from polarimetric studies, where some continuum is seen to escape without 
passing through dust (e.g. Hines \& Wills 1995).   Powerful QSOs, especially
powerful radio sources, are able to break through the dust cocoon more 
rapidly, and this is why we do not see many strong radio-loud quasars in
cocoon phase (Canalizo \& Stockton 2001; Gregg et al. 2000).

\section{Biases and Future Work}
What keeps the evidence from having more weight is 
the fact that the sample of lo-BAL QSOs at $z < 0.4$ is incomplete and 
may be significantly biased.   Only a fraction of low-redshift QSOs have been
observed in the UV (where the BAL features are evident), and these 
observations have different biases.   For example, Turnshek et al. (1997)
conducted an HST FOS survey of low-z BALs in a sample of QSOs with weak
[O\,III] (and strong Fe\,II).   These two properties have an unusually high
incidence in young IR-loud QSOs (Canalizo \& Stockton 2001). The Turnshek
et al. sample may then be biased towards young objects. 
We are therefore obtaining further spectroscopic and high resolution 
imaging observations of the host galaxies of BAL QSOs at somewhat higher 
redshifts (for which the BAL features appear in the optical).

The possibility that the lo-BAL phenomenon represents a short
phase in the early life of QSOs is of great interest because it could
potentially provide a method to answer one of the fundamental questions
regarding QSOs, namely, how long does the QSO activity last?
Determining the ages of starbursts that are related to the fueling of QSOs
in a sample of objects could place limits on the duration of the BAL
phase.   This, along with an estimate of the fraction of QSOs go through a 
lo-BAL phase (as seen from our line of sight) 
would place upper limits on the mean lifetime of the QSO activity.


\acknowledgments
Part of this work was performed under the auspices of the U.S.\ Department 
of Energy, National Nuclear Security Administration by the University of 
California, Lawrence Livermore National Laboratory under contract 
No.\ W-7405-ENG-48, and was also partially 
supported by NSF under grant AST 95-29078.  Based on observations with 
the NASA/ESA Hubble Space Telescope, obtained from the data archive at 
the Space Telescope Science Institute, which is operated by the Association 
of Universities for Research in Astronomy, Inc. under NASA contract 
No.\ NAS5-26555.

\end{document}